\documentclass[12pt]{article}
\usepackage{amsmath,amssymb}
\usepackage{graphicx,color}
\numberwithin{equation}{section}
\usepackage{epsfig}
\usepackage{cite}
\usepackage{bm}
\usepackage{dcolumn}
\newcommand{\nn}{\nonumber}
\newcommand{\be}{\begin{equation}}
\newcommand{\bea}{\begin{eqnarray}}
\newcommand{\eea}{\end{eqnarray}}
\newcommand{\ba}{\begin{array}}
\newcommand{\ea}{\end{array}}
\newcommand{\ee}{\end{equation}}

\expandafter\ifx\csname mathbbm\endcsname\relax

\else

\fi \textheight 22cm \textwidth 15cm \topmargin 1mm \oddsidemargin
5mm \evensidemargin 5mm

\begin{document}
\begin{titlepage}
\hfill \vbox{
    \halign{#\hfil         \cr
          % IPM// \cr
                      } % end of \halign
      }  % end of \vbox
\vspace*{20mm}
\begin{center}
{\Large {\bf Noncritical Holographic QCD in External Electric Field  }\\
}

\vspace*{15mm} \vspace*{1mm} {Ali Davody}

 \vspace*{1cm}

{\it ${}^b$ School of Particles and Accelerators, Institute for Research in Fundamental Sciences (IPM)\\
P.O. Box 19395-5531, Tehran, Iran \\ }

\vspace*{.4cm} {davody@ipm.ir}

\vspace*{2cm}
%%\maketitle
\end{center}

\begin{abstract}
We investigate  behavior of a noncritical model in external
electric field and explore its phase structure in the quenched
approximation $N_f\ll N_c$. We compute the conductivity of QCD
plasma in this model and compare it  with the predictions
of Sakai-Sugimoto model, D3-D7 system and the  lattice simulation. We
find that, while  the behavior of conductivity in noncritical model as a
function of temperature and baryon density is similar to those
 of D3-D7 system, the phase diagram of noncritical
model resembles the phase diagram of Sakai-Sugimoto model.

\end{abstract}
\end{titlepage}

\section{introduction}
The AdS/CFT correspondence is a duality between the strongly coupled
conformal field theories  and string theory in  a higher-dimensional
Anti-de-Sitter  space-time \cite{Maldacena1997re}. The
generalization of AdS/CFT correspondence to more realistic gauge
theories like QCD, provides new insights to  understanding the
dynamical non-perturbative effects in  QCD,  such as  confinement,
chiral symmetry breaking, color superconductivity and so on.

Dual gauge theories arising from  brane constructions in
ten-dimensional critical string theory are supersymmetric. In order
to break supersymmetry, one may   compactify  the supersymmetric
guage theory on a circle of radius $R$ and impose anti-periodic
boundary conditions for fermions around the circle. The resulting
effective theory at low energy compared to the Kaluza-Klein mass
scale, $M_{KK}\sim \frac{1}{R}$, is pure QCD without fundamental
matter. Fundamental matter  may be incorporated  by  adding flavor branes
\cite{Karch:2002sh}.
%One then add flavor branes to incorporate fundamental matter
%multiplets  \cite{Karch:2002sh}.
In principle, it is possible to  extract the
physics of strongly correlated QCD by using these holographic
models. The main obstacle of this approach, however, is that
resulting holographic QCD contains undesired Kaluza-Klein
 modes with the mass as the same order of  hadrons and glueballs. One way  to overcome this problem
  is to consider  brane backgrounds in noncritical string theory \cite{Klebanov:2004ya}. Since in this case
holographic backgrounds live in lower dimensions, the problem of
extra KK modes is  more  tractable .

A noncritical holographic model of QCD has been introduced in
\cite{Kuperstein:2004yk}, where a stack of $N_c$ D4 branes in
six-dimensional noncritical
 string theory play the role of  color branes. Fundamental matters may  be added by inserting $N_f$ D4 branes
and $\overline{\text{D4}}$ breanes in this
background\cite{Casero:2005se}. Further investigations
\cite{Casero:2005se,Kuperstein:2004yf} have shown that this model
captures correctly many properties of low energy QCD, like mass
spectrum of mesons, area low behavior for the Wilson loop,
glueball mass spectra, etc. In particular, a holographic
calculation of  Wilson line has been performed recently in \cite{Kol:2010fq} for
near extremal D3 branes, D4 branes, non-critical near extremal AdS$_6$
model and  Klebanov-Strassler model. It has been shown  that the noncritical background
admits a reasonable fit to lattice results (Actually the
Klebanov-Strassler background exhibits the best fit to lattice
results, but noncritical model has a asymptotic AdS$_6$ metric, see
 for details \cite{Kol:2010fq}). The thermal phases of the model have  been
investigated in \cite{Mazu:2007tp}, where  it was  shown that
there is a first order confined/deconfined phase transition at
$T_c=\frac{1}{2\pi R}$. In similar to  Sakai-Sugimoto model,
chiral symmetry restores above a critical temperature $T_{\chi
SB}=\frac{0.169}{L}$, where $L$ is the separation of  flavors
branes at infinity.

It is also interesting to
investigate the behavior of QCD matter in presence of other
fundamental interactions. Indeed  the physics of some astrophysical
phenomena like quark stars and  notron stars is related
to the behavior of thermal QCD in external electromagnetic field. For example
 conductivity of QCD matter encodes the timescale for
expulsion of magnetic flux lines from the core of quark star (see for
example \cite{Ouyed:2003ge}). Since QCD is strongly coupled  at
these circumstance, it is natural to use AdS/CFT  and   extracting
the properties of QCD at these high densities and temperatures.
Using gauge/garvity duality,  response of Sakai-Sugimoto model to
external electric and magnetic field has been studied in
\cite{Bergman:2008sg,Kim:2008zn,Johnson:2009ev,Johnson:2008vna}, behavior of massive ${\cal{N}} =2$
hypermultiplets in an ${\cal{N}} =4$ SYM plasma (D3-D7 system) in presence of background electric and 
magnetic field  was explored in\cite{Karch:2007pd,Albash:2007bq,Albash:2007bk} and the  effect of magnetic field on
dynamics of noncritical holographic model has been considered
in\cite{Cui:2009ux}.
%was explored in ?

%
%On the other hand our knowledge of the behavior thermal QCD in external electric and
%magnetic field provide a window for understanding some problems like ..........  Since QCD is strong coupled in this situations
%ads cft  ..........................

Motivated by these discussions, in this paper, we consider a
noncritical holographic model of QCD, and examine its behavior in
the presence of external electric field. We  compare the results of
this model with the predictions of Sakai-Sugimoto model
\cite{Bergman:2008sg}, D3-D7 system \cite{Karch:2007pd}, and a lattice
simulation \cite{Gupta:2003zh}. We find that the behavior of
conductivity in noncritical model as a function of temperature and
baryon density is similar to the results of D3-D7 system. In
weak field regime, where lattice simulation is accessible,
noncritical holographic model predicts a finite conductivity for
deconfined QCD matter  which is  linear in temperature, in good
agreement with lattice QCD result. Also, the phase diagram of
noncritical model resembles the phase diagram of Sakai-Sugimoto
model. In particular we find that electric field reduces
chiral-symmetry restoration temperature.

This paper is organized as follows: In section 2 we briefly review
the noncritical model. In section 3 we analyze  the dynamics of
flavor branes in presence of electric field and extract the phase
diagram of the model. We check our result by making use of Kubo
formula in section 4 and section 5 is devoted to brief summary and
conclusions.

\section{Review of the model}
In this section we briefly review the noncritical model
of\cite{Kuperstein:2004yk}. The model is based on
D4/D4-$\overline{\text{D4}}$  brane system, where $N_c$ D4-branes
compactified  on $S^1$ with radius $R$ and $N_f$
D4-$\overline{\text{D4}}$ flavor branes are transverse to the $S^1$.
By imposing periodic boundary condition on    bosonic field  and
antiperiodic boundary condition  on   fermionic field  along the
circle $S^1$, supersymmetry is broken and one obtains QCD spectrum
at the energies below the Kaluza-Klein scale. The  brane
configuration of the system is

\begin{equation}
\begin{tabular}{ccccccccccc}
& $t$ & $x_1$ & $x_2$ & $x_3$ & $x_4$ & $x_5$  \\ \hline D4 & $\times$ &
$\times$ & $\times$ & $\times$ & $\times$ &   \\
D4-$\overline{\text{D4}}$ & $\times$ & $\times$ & $\times$ &
$\times$
& $$ & $\times$ &  \\
\end{tabular} \nonumber
\end{equation}

where $x_4$ is the coordinate of $S^1$. At  zero temperature  the
6-dimensional background metric  is given by

\bea\label{lowbackground}
ds_6^2=\bigg(\frac{u}{R_{AdS}}\bigg)^2(-dt^2+dx^idx^i+f(u)\,dx_4^2)+\bigg(\frac{R_{AdS}}{u}\bigg)^2 \frac{du^2}{f(u)}\\\nn
f(u)=1-\big{(}\frac{u_{\Lambda}}{u}\big{)}^5, \;\;\;\;\; \;\;\;\;\;\;\;\;\;\; R_{AdS}=\sqrt{\frac{15}{2}}\;l_s
\eea

There is also a constant dilaton and a 6-form field strength

\bea &F_{(6)}&=- Q_c \bigg(\frac{u}{R_{AdS}}\bigg)^4 dx_0\wedge
dx_1\wedge dx_2\wedge dx_3\wedge dx_4\wedge du \nn\\
&e^\phi&=\frac{2\sqrt2}{\sqrt3 Q_c} \nn \eea

The space in $(x_4,u)$ plane looks like a cigar where tip located at $u=u_{\Lambda}$, and  to avoid a conical singularity the periodicity of
$x_4$ should satisfies

\be\label{periodx4} x_4\sim x_4 +2\pi R=x_4+\frac{4\pi
R_{AdS}^2}{5u_{\Lambda}} \ee

Kaluza-Klein energy scale  is $M_{KK}=\frac{1}{R}=\frac{5u_{\Lambda}}{2
R_{AdS}^2}$, below this scale the effective theory is QCD$_4$. At
nonzero temperature there are two solutions with the same boundary
condition \cite{Mazu:2007tp}. The one which dominates at low temperature is given by
(\ref{lowbackground}) where  the periodicity of euclidian time is
arbitrary, $t_E\sim t_E+\beta$, and periodicity of $x_4$ is given by
(\ref{periodx4}). The flavor branes form a U embedding
configuration, so this  background corresponds to the confined phase
with broken chiral symmetry.

By increasing temperature a confinement/deconfinement phase transition occurs
 at $T_c=\frac{1}{2\pi R}$, and the background for  $T_c<T$ is represented by
\bea
ds^2=\big{(}\frac{u}{R_{AdS}}\big{)}^2(f(u) dt_E^2+dx_i dx_i+dx_4^2)+\big{(}\frac{R_{AdS}}{u}\big{)}^2 \frac{1}{f(u)}du^2\\\nn
f(u)=1-\big{(}\frac{u_{T}}{u}\big{)}^5, \;\;\;\;\; \;\;\;\;\;\;\;\;\;\; t_E\sim t_E+\frac{4 \pi R_{AdS}^2}{5 u_T}
\eea

In this case  the periodicity of $x_4$ circle is arbitrary and
temperature is given by $T=\frac{5 u_{T}}{4\pi R_{AdS}^2}$. This
background allows  two embeddings for
D4-$\overline{\text{D4}}$ flavor branes:  U embedding which is
preferred configuration for $T_c<T<T_{\chi SB}$ and parallel
embedding which dominates for $T>T_{\chi SB}$, where $T_{\chi
SB}=\frac{0.169}{L}$. Thus chiral symmetry restores at temperatures
above $T_{\chi SB}$.

%&&&&&&&&&&&&&&&&&&&&&&&&&&&&&&&&&&&&&&&&&&&&&&&& Section 2 &&&&&&&&&&&&&&&&&&&&&&&&&&&&&&&&&&&&&&&&&&&&&&&&&&&&&&&&&&&&&&&&&&&&&
\section{Adding $U(1)$ gauge field}
In order to accommodate an external electric field, following
\cite{Karch:2007pd}, we turn on a $U(1)$ gauge field on flavor
branes

\be\label{gauge} A_x(t_E,u)=-iEt_E+h(u) \ee

where time dependence part of gauge field describes a static
electric field on the boundary and $u$ dependence part encodes the
response current. In this section we analyze various phases of
noncritical holographic QCD in external electric field by making use
the method of \cite{Karch:2007pd} and \cite{Bergman:2008sg}.

\subsection{Deconfined Phase}

Let us start with investigating the  effect of an external electric field on  high temperature ($T>T_c$) phase of QCD.
By using $\xi=(t_E,\overrightarrow{x},u)$ for parametrization of  D4-$\overline{\text{D4}}$ flavor branes,
the  induced metric on flavor branes
is given by

\bea ds^2=\big{(}\frac{u}{R_{AdS}}\big{)}^2(f(u) dt_E^2+dx_i
dx_i)+\bigg{(}(\frac{u}{R_{AdS}})^2 {x_4'}^2+(\frac{R_{AdS}}{u})^2
\frac{1}{f(u)}\bigg{)}du^2 \eea

DBI action in presence of background gauge field (\ref{gauge})
takes the following form \footnote{Throughout this paper we work
with action density }
%
%Now consider the  DBI action   for flavor  branes
%\be
%S_{D4-D\overline{4}}= 2 N_f T_4 \int d^5\xi e^{-\phi} \sqrt{det(g_{ab}+2\pi \alpha ' F_{ab})}
%\ee
%
%
%
%
%the DBI action takes the following form
\bea
S=\frac{{\cal{N}}}{R_{AdS}^5}\int du u^5 \sqrt{(f(u){x_4'}^2
 + \frac{R_{AdS}^4}{u^4})(1-\frac{e^2 R_{AdS}^4}{u^4 f(u)})+ \frac{f(u) R_{AdS}^4 }{u^4} {a_x'}^2}\\
\eea

where ${\cal{N}}=2 N_f T_4 e^{-\phi}, \; e=2\pi\alpha ' E$,  and
$\;a_x'=2\pi\alpha ' A_x'$. By making use of the first integrals of motion one can find the asymptotic forms  of gauge field and $x_4'$ at large
$u$
\bea
a_x'\sim  \frac{j}{u^3},\;\;\;\;\;\;\;\;\;\;\;x_4'\sim\frac{c}{u^7}\\
%A_x\sim_{u\longrightarrow \infty} -iet_E -\frac{\alpha}{2} \frac{1}{u^2}
\eea

where $j$ and $c$ are identified with   the current and chiral condensate in dual gauge theory.
By doing holographic renormalization one can show that the physical current  on the boundary is

\be
J^{x}=\frac{2\pi \alpha'{\cal{N}}}{R_{AdS}^3}j
\ee

By writing the action in terms of current, $j$, we have
\be\label{uaction}
S=\frac{{\cal{N}}}{R^5}\int du u^5 \sqrt{(f(u){x_4'}^2 +
\frac{R_{AdS}^4}{u^4})(f(u)-\frac{e^2 R_{AdS}^4}{u^4 })( f(u)-\frac{j^2 }{u^6} )^{-1}}
\ee

As we have mentioned in the previous section, deconfined  background
possesses  two embeddings, U embedding and parallel embedding. Our
aim is to determine current $j$ for a given external
electric field in these  embeddings and   then extracting  conductivity of  these configurations.

First consider the U embedding. In the case of vanishing response
current, $j=0$, the solution takes the following form

\bea
 x_4'(u)=\frac{R_{AdS}^2}{u^2\sqrt f}\bigg{[}\frac{u^{10}(f(u)-
 \frac{e^2 R_{AdS}^4}{u^4 })}{u_0^{10}(f(u_0)-\frac{e^2 R_{AdS}^4}{u_0^4 })}-1\bigg{]}^{\frac{-1}{2}}
\eea
where $u_0$ is the turning point,  $x_4'(u_0)=0$. By inserting
the above solution into the action  we arrive at

\be
S_U=\frac{{\cal{N}}}{ R_{AdS}^3 }\int_{u_0} ^{\infty}du \, u^3
 \sqrt{1-\frac{e^2 R_{AdS}^4}{u^4 f(u)}}\bigg{[}1- \frac{u_0^{10}(f(u_0)-
 \frac{e^2 R_{AdS}^4}{u_0^4 })}{u^{10}(f(u)-\frac{e^2 R_{AdS}^4}{u^4 })}\bigg{]}^{\frac{-1}{2}}
\ee

 As long as
 $e_0^2\leq \frac{1}{R_{AdS}^4}u_0^4 f(u_0)$, the action is real and embedding is acceptable.
 Since by turning on the current,  action increased, the fevered configuration is a U embedding with vanishing current, $j=0$ \cite{Bergman:2008sg}.
 In dual QCD this means  that the deconfined chiral-symmetry breaking phase is an insulator.

A natural question is that what happens for $e>e_0$?.  Figure 1 depicts
$L$ as a function of  $c$ for different values of electric field. As
it is evident from this figure,   $L$ is a decreasing function of $e$,
so there is a maximal value of $e$ at fixed values of $L$ and $T$, such that
above which there are no U embedding solution. Thus we expect that
the favorite solution in this regime becomes parallel embedding and
there should be a phase transition from chiral-broken phase to
chiral symmetric phase by increasing electric field \cite{Bergman:2008sg}. We come back to
this issue after discussing the behavior of   parallel embeding in
electric field.

\begin{figure}[!h]
\begin{center}
\includegraphics[width=.6\textwidth]{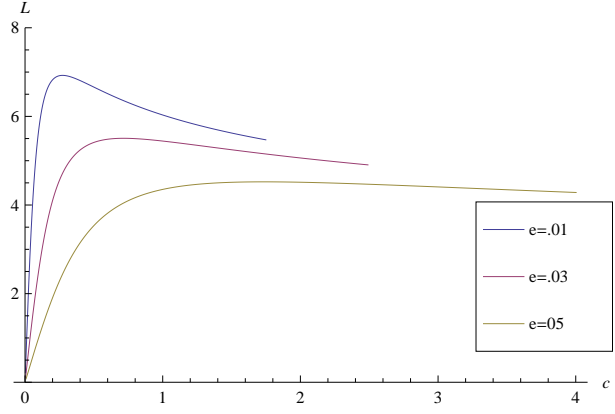}
 \end{center} \caption{$L$ as a function of $c$ in deconfined phase for different values of electric field
 %(Green color is corresponding to $e=$, blue $e=$ and red $e=$ )
 }\label{lcdeconfined}
\end{figure}

In the parallel embedding the DBI  action becomes
\be
S_{||}=\frac{{\cal{N}}}{ R_{AdS}^3 }\int_{u_T} ^{\infty}du \,
u^3 \sqrt{\frac {f(u)-\frac{e^2 R_{AdS}^4}{u^4}}{ f(u)-\frac{j^2}{u^6}}}
\ee

From this expression,  it is clear that action becomes complex
somewhere unless a nonzero current  being turned on. The magnitude
of current is given by \be j=R_{AdS}^2\, e\, u_c\ \ee where $u_c$ is
the root of numerator

\be\label{root} f(u_c)-\frac{e^2 R_{AdS}^4}{u_c^4}=0 \ee

Therefore the chiral-symmetry restoration  phase behaves like a conductor by following conductivity

\bea\label{connon} \sigma_{non-critical}=\frac{J^{x}}{E}=\frac{(2\pi
\alpha')^2 {\cal{N}}}{R_{AdS}^3}\frac{j}{e}=\frac{(2\pi \alpha')^2
{\cal{N}}}{R_{AdS}}\, u_c(e,T) \eea

There is no algebraic solution for (\ref{root}), however, we can
study the weak filed and strong field behavior of conductivity

\bea
\sigma_{non-critical}=\left\{
  \begin{array}{ll}
    \sqrt{\frac{12\pi}{5}}\,(2\pi \alpha')^{\frac{5}{2}} {\cal{N}}\;\,T
     & \hbox{ $E\ll \frac{12\pi}{5}\,T^2$ }
    \\ \\
   (2\pi \alpha')^{\frac{5}{2}} {\cal{N}}\,\,E^{\frac{1}{2}} & \hbox{$E\gg \frac{12\pi}{5}\,T^2$}
  \end{array}
\right.
\eea

It is interesting to compare this results with the predictions of
Sakai-Sugimoto  model, $\sigma_{S-S}$,
and D3-D7 system, $\sigma_{D3-D7}$. The weak and high field
behavior of conductivity in these models are given by \cite{Bergman:2008sg,Karch:2007pd}

\bea\nn \sigma_{S-S}=\left\{
  \begin{array}{ll}
    \frac{N_f N_c }{27\pi} \lambda_5\; T^2
    & \hbox{ $E\ll \frac{8\pi^2}{27}\lambda_5 T^3$ }
    \\ \\
  \frac{N_f N_c }{12\pi^{\frac{7}{3}}} \lambda_5^{\frac{1}{3}}\; E^{\frac{2}{3}}
    & \hbox{ $E\gg \frac{8\pi^2}{27}\lambda_5 T^3$ }
  \end{array}
\right. \sigma_{D3-D7}=\left\{
  \begin{array}{ll}
    \frac{N_f N_c }{4\pi}\;  T
     & \hbox{$ E\ll \frac{\pi}{2}\sqrt{\lambda_4}T^2$ }
    \\ \\
  \frac{N_f N_c }{(2\pi)^{\frac{3}{2}}\lambda_4^{\frac{1}{4}}}\;E^{\frac{1}{2}}
   & \hbox{$E\gg \frac{\pi}{2}\sqrt{\lambda_4}T^2$}
  \end{array}
\right.
\eea

%\begin{figure}[!h]
%  % Requires \usepackage{graphicx}
%\begin{center}
%\includegraphics[width=.5\textwidth]{maxwellconstruction.eps}
% \end{center} \caption{Maxwell construction }\label{maxwell}
%\end{figure}
where $\lambda_5$ and $\lambda_4$ stands for  five and four
dimensional t' Hooft coupling. As it is clear  from above
expressions, the behaviour of conductivity as a function of
temperature and electric field in noncritical model is  similar to
the D3-D7 system.

Now, let us consider to the phase structure of deconfined
background. Since external electric field bring the system out of
equilibrium, as argued in \cite{Bergman:2008sg}, to determine which
phase is favorite, one can use a Maxwell Like construction for order
parameter $c=\frac{\partial F_e}{\partial L}\mid_{e,T}$.
% The dependence of $L$ on $c$ in $U$ embedding is shown in fig. 2.
According to Maxwell construction, for a fixed value of $L$ and $T$,
the transition occurs at the value of $e$ such that two areas A and B (see figure 2(a)) become equal. By changing
 electric field we find the phase diagram as  figure 2(b), which is, remarkably, similar to those of
 Sakai-Sugimoto model  \cite{Bergman:2008sg}. According to this phase
 diagram, we observe that the critical temperature decreases with increasing
  external electric field, as we expected from the polarization
 effect of electric field. Also  at zero electric field we
 get the result of \cite{Mazu:2007tp} for chiral symmetry restoration temperature, $T_{\chi
 B}=\frac{0.169}{L}$.

%
%\begin{figure}
%  % Requires \usepackage{graphicx}
%\begin{center}
%\includegraphics[width=.5\textwidth]{deconfinedphasediagramfinal.eps}
% \end{center} \caption{phase diagram fort $L=1$ in presence of external electric field in deconfined background}\label{phasedeconfined}
%\end{figure}

\begin{figure}[htbp]
\begin{center}
\begin{tabular}{ccc}
\epsfig{file=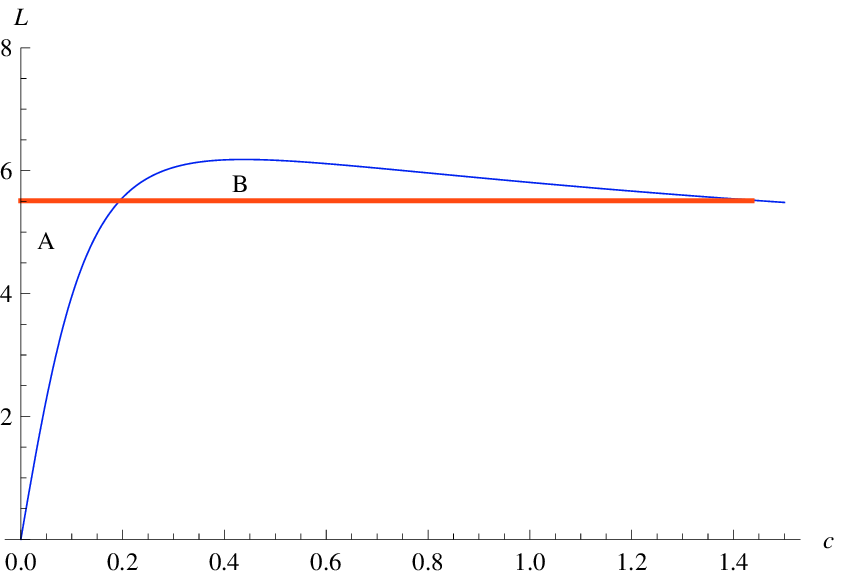,scale=0.7} \;\;\;\;\; &
\epsfig{file=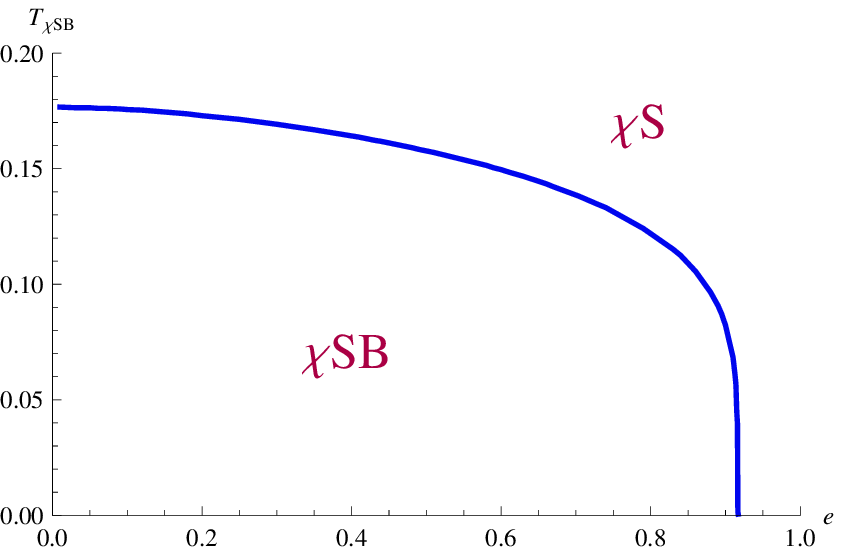,scale=0.7} \;\;\;\;\; &
\end{tabular}
\caption{{
(a) Maxwell construction (b) phase diagram fort $L=1$ in presence of
 external electric field in deconfined background}}
\label{embeddings}
\end{center}
\end{figure}

%%%%%%%%%%%%%%%%%%%%%%%%%%%%%%%%%%%%%%%%%%%%%%%%%%%%%%%%%%%%%%%%%%%%%%%%%%%%%%%
%%%%%%%%%%%%%%%%%%%%%%%%%%%%%%%%%%%%%%%%%%%%%%%%%%%%%%%%%%%%%%%%%%%%%%%%%%%%%%%%
\subsubsection{finite density}

In this subsection,  we  generalize our analysis in the parallel
embedding for finite baryon density by turning on a nontrivial
zero-component of the gauge field $a_0(u)$. The resulting  DBI action has the form

\bea\nn S=\frac{{\cal{N}}}{R_{AdS}^5}\int du u^5
\sqrt{(f(u){x_4'}^2 + \frac{R_{AdS}^4}{u^4})(1-\frac{e^2
R_{AdS}^4}{u^4 f(u)})+ \frac{f(u) R_{AdS}^4 }{u^4} {a_x'}^2-\frac{
R_{AdS}^4 }{u^4} {a_0'}^2} \eea

Solving the equation of motion one finds the asymptotic behavior of
gauge fields as

\be
a_x\simeq constant-\frac{j}{2  u^2} \;\;\;\;\;\;\;\;\;\;\;\;\;
a_0\simeq constant+\frac{d}{2  u^2}
\ee

where $j$ and $d$  are related to the physical current and charge
density via

\be
J=\frac{(2\pi\alpha'){\cal{N}}}{R_{AdS}^3}j\;\;\;\;\;\;\;\; D=\frac{(2\pi\alpha'){\cal{N}}}{R_{AdS}^3}d
\ee
 Writing the action  in terms of current and charge density we arrive at

\be
S=\frac{{\cal{N}}}{R_{AdS}^5}\int du   u^3 \sqrt{\frac{1-\frac{e^2 R_{AdS}^4}{u^4 f(u)}}{1-\frac{j^2-d^2 f(u)}{u^6 f(u)}}}
\ee

Reality of action implies a nonzero current

\be j=R_{AdS}^2 \sqrt{u_c^2+\frac{d^2}{u_c^4}} \;\;e \ee

from which conductivity at finite charge density can be read as

\be\label{conden}
\sigma=\frac{J}{E}=\frac{(2\pi\alpha')^2{\cal{N}}}{R_{AdS}}\sqrt{u_c^2+\frac{d^2}{u_c^4}}
\ee

At  high density approximation it becomes

\bea
\sigma_{noncritical}=\frac{25\alpha'^2{\cal{N}}}{4R_{AdS}^5}\frac{d}{T^2}=\frac{5}{12\pi}\frac{D}{T^2}
\eea

Let us compare this  result with that one  obtained in
Sakai-Sugimoto model and D3-D7 system

\bea
\sigma_{Sakai-Sugimoto}=\frac{27}{8\pi^2\lambda_5}\frac{D}{T^3} \;\;\;\;\;\;\;\;\;\;\;\;\;\;\;\;\;\;\; \sigma_{D3-D7}=\frac{2}{\pi
\sqrt{\lambda_4}}\frac{D}{T^2}
\eea

Again we observe that charge density dependence of conductivity in
 noncritical model and  D3-D7 system is the same.

%the behavior of conductivity  as a function of charge density and temperature,
%in noncritical model and  D3-D7 system is the same.

%%%%%%%%%%%%%%%%%%%%%%%%%%%%%%%%%%%%%%%%%%%%%%%%%%%%%%%%%%%%%%%%%%%%%%%%%%%%%%%%%%%%%%%%%%%%%%%%%%%%%%
%%%%%%%%%%%%%%%%%%%%%%%%%%%%%%%%%%%%%%%%%%%%%%%%%%%%%%%%%%%%%%%%%%%%%%%%%%%%%%%%%%%%%%%%%%%%%%%%%%%%%%
\subsection{confined phase} In this section we will study the
response of confined phase to external electric field. This
background dominates for $T<T_c$ and in the case of absence external
field, the  only allowed embedding for flavor branes is U embedding . To
proceed we start with the DBI action for flavor branes

\bea
S=\frac{{\cal{N}}}{R_{AdS}^5}\int du u^5 \sqrt{(f(u){x_4'}^2 + \frac{R_{AdS}^4}{u^4})(1-\frac{e^2 R_{AdS}^4}{u^4 })+
\frac{ R_{AdS}^4 }{u^4} {a_x'}^2}
\eea

First consider the $U$ embedding with $j=0$. The equations of motions gives

\bea
 x_4'(u)=\frac{R_{AdS}^2}{u^2 f(u)}\bigg{[}\frac{u^{10}f(u)(1-
 \frac{e^2 R_{AdS}^4}{u^4 })}{u_0^{10}f(u_0)(1-\frac{e^2 R_{AdS}^4}{u_0^4 })}-1\bigg{]}^{\frac{-1}{2}}
\eea

with the following asymptotic behavior near the boundary

\be
x_4'\simeq\frac{c}{u^7}\;\;\;\;\;\;\;\;\;
\ee

where

\be
c=R_{AdS}^2 u_0^5 \sqrt{f(u_0)(1-\frac{e^2 R_{AdS}^4}{u_0^4})}
\ee

In order to have a physical solution, it is required
$u_0>e^{\frac{1}{2}}R_{AdS}$. Since the range of coordinate $u$ is
$(u_{KK},\infty)$, if $e<\frac{u_{KK}^2}{R_{AdS}^2}$, $u_0$ can
interpolates between $u_{KK}$ and $\infty$. Figure 3(a) shows   the
behavior of $L$ in terms of $c$, in this case. On the other hand,
for $e>\frac{u_{KK}^2}{R_{AdS}^2}$, the range of $u_0$ would be
$(e^{\frac{1}{2}}R_{AdS},\infty)$ and the plot of $L(c)$ is like
deconfind  background (figure 3(b)). Then  for
$e>\frac{u_{KK}^2}{R_{AdS}^2}$, there is a maximal $L$ for a given
$e$. Since by increasing $e$ the maximal of $L$ decreases, there is
a maximal $e$ at any fixed value of $L$, such that  for $e>e_{max}$, U
embedding is not a valid solution. As discussed in
\cite{Bergman:2008sg} the  solution for $e>e_{max}$ is a V
shape embedding, where flavor branes follow parallel radial
geodesics. Thus  this solution satisfies $x_4'(u)=0$ except at the tip.
DBI action for D4-$\overline{\text{D4}}$ branes in this background
is as follows

%\begin{figure}
%  % Requires \usepackage{graphicx}
%\begin{center}
%\includegraphics[width=.5\textwidth]{l(c)adeconfined.eps}
% \end{center} \caption{$L$ as a function of c in confined phase }\label{phasedeconfined}
%\end{figure}
%
%\begin{figure}
%  % Requires \usepackage{graphicx}
%\begin{center}
%\includegraphics[width=.5\textwidth]{l(c)bdeconfined.eps}
% \end{center} \caption{$L$ as a function of c in confined phase }\label{phasedeconfined}
%\end{figure}

\begin{figure}[htbp]
\begin{center}
\begin{tabular}{ccc}
\epsfig{file=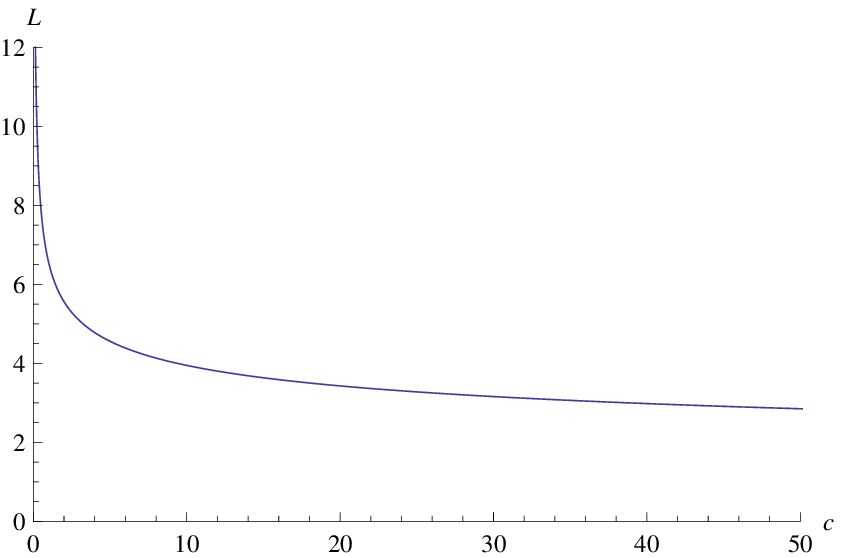,scale=0.8} \;\;\;\;\; &
\epsfig{file=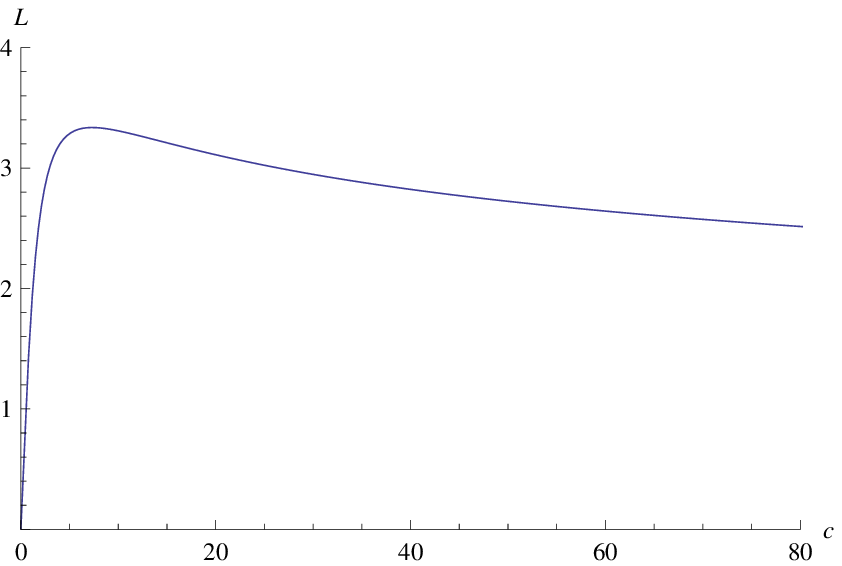,scale=0.8} \;\;\;\;\; &
\end{tabular}
\caption{{$L$ as a function of c in confined phase
(a) $e<\frac{u_{KK}^2}{R_{AdS}^2}$ (b) $e>\frac{u_{KK}^2}{R_{AdS}^2}$ }}
\label{embeddings}
\end{center}
\end{figure}

%\be
%S_U=\frac{{\cal{N}}}{ R_{AdS}^3 }\int_{u_0} ^{\infty}du \, \frac{u^3}{\sqrt f(u)}
% \sqrt{1-\frac{e^2 R_{AdS}^4}{u^4 }}\bigg{[}1- \frac{u_0^{10}f(u_0)(1-
% \frac{e^2 R_{AdS}^4}{u_0^4 })}{u^{10}f(u)(1-\frac{e^2 R_{AdS}^4}{u^4 })}\bigg{]}^{\frac{-1}{2}}
%\ee

\be
S_V=\frac{{\cal{N}}}{ R_{AdS}^3 }\int_{u_T} ^{\infty}du \,
\frac{u^3}{\sqrt{f(u)}} \sqrt{\frac {1-\frac{e^2 R_{AdS}^4}{u^4}}{ 1-\frac{j^2}{u^6}}}
\ee

From which one can read the conductivity of $V$ embedding
\be
\sigma=(2\pi\alpha')^{\frac{5}{2}} {\cal{N}} \;E^{\frac{1}{2}}
\ee

Therefore  the confined phase exhibits two phases in external
electric field. $U$ embedding which describes an insulating phase
and $V$ embedding with a finite electric conductivity. In order to
explore the phase diagram, we apply  the same method used for
deconfined phase. The resulting phase diagrams for fixed value of
$u_{KK}$ and $L$ is shown in Figure 4, which are similar to those of Sakai-Sugimoto model \cite{Bergman:2008sg}.
 From this diagram we observe that U embedding is dominate background for
$e<\frac{u_{KK}^2}{R_{AdS}^2}$, disregard how much $L$ is, as we
expected.
\begin{figure}[htbp]
\begin{center}
\begin{tabular}{ccc}
\epsfig{file=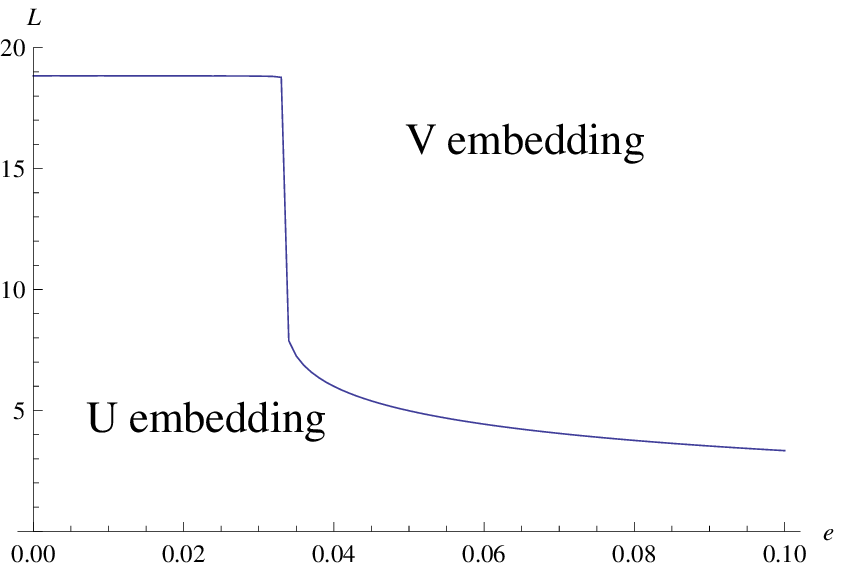,scale=0.8} \;\;\;\;\; &
\epsfig{file=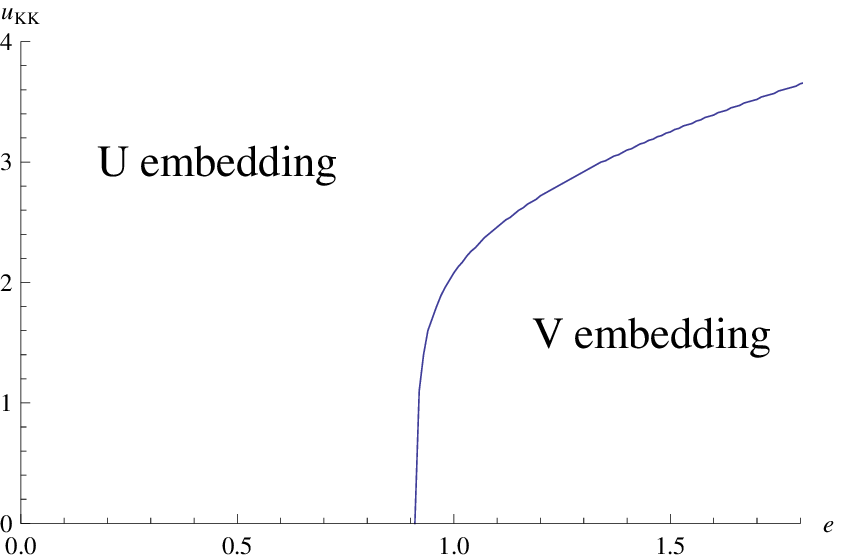,scale=0.8} \;\;\;\;\; &
\end{tabular}
\caption{{ phase diagram in confined phase, (left) $u_{KK}=.5$, (right) L=1  }}
\end{center}
\end{figure}

%\begin{figure}
%  % Requires \usepackage{graphicx}
%\begin{center}
%\includegraphics[width=.5\textwidth]{noncritical-confined-phasediagrama.eps}
% \end{center} \caption{ phase diagram in confined phase  }\label{phasedeconfined}
%\end{figure}
%\begin{figure}
%  % Requires \usepackage{graphicx}
%\begin{center}
%\includegraphics[width=.5\textwidth]{noncritical-confined-phasediagramb.eps}
% \end{center} \caption{ phase diagram in confined phase  }\label{phasedeconfined}
%\end{figure}

%%%%%%%%%%%%%%%%%%%%%%%%%%%%%%%%%%%%%%%%%%%%%%%%%%%%%%%%%%%%%%%%%%%%%%%%%%%%%%%%%%%%%%%%%%%%%%%%%%%%%%%%%%%%%%
\subsection{electric susceptibility}
The response of an insulating phase to an external electric field
can be measured by electric susceptibility, which defined by \be
\chi_e=-\frac{\partial^2 F_e}{\partial e^2} \ee

Since there is an insulating phase in both deconfined and confined
background, it is interesting  to use holographic description to
calculate the susceptibility in the dual QCD. Indeed holographic
free energy is divergent and one has to add a counterterm to find
finite results. We apply regularization of \cite{Bergman:2008sg}

\be \chi_e=-\frac{\partial^2 F_e}{\partial e^2}+\frac{\partial^2
F_e}{\partial e^2}\mid_{e=0} \ee

Using this expression, we find electric susceptibility in deconfined
and confined phase according to figure 5.
\begin{figure}[htbp]
\begin{center}
\begin{tabular}{ccc}
\epsfig{file=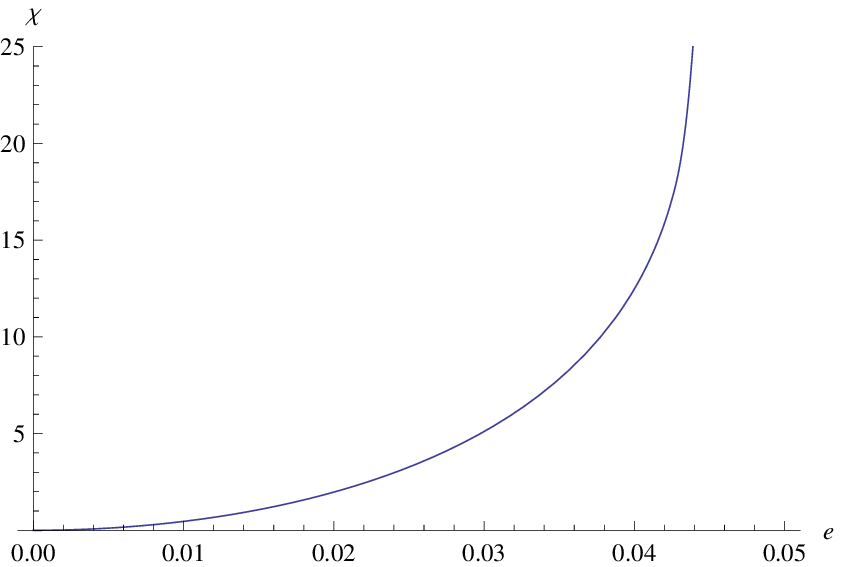,scale=0.8} \;\;\;\;\; &
\epsfig{file=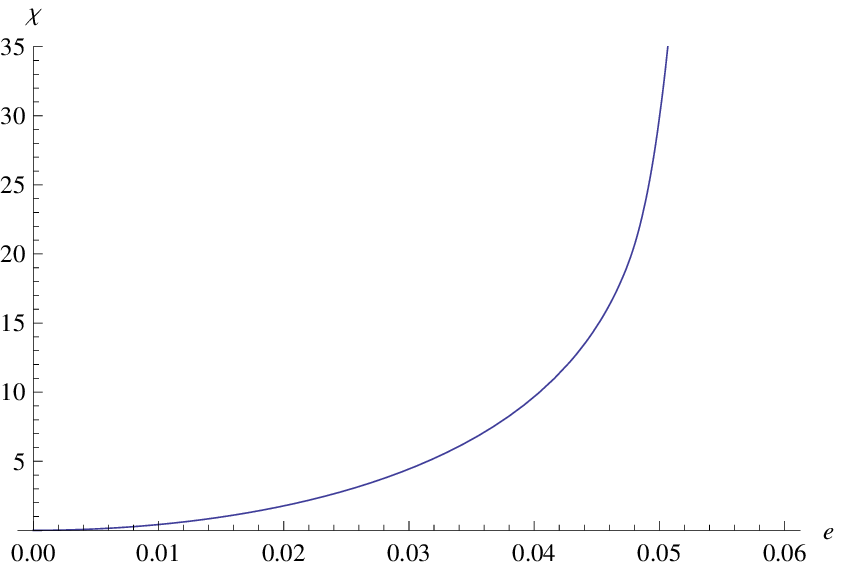,scale=0.8} \;\;\;\;\; &
\end{tabular}
\caption{{  susceptibility in deconfined phase (left) and confined phase (right)}}
\label{embeddings}
\end{center}
\end{figure}

%\begin{figure}
%  % Requires \usepackage{graphicx}
%\begin{center}
%\includegraphics[width=.5\textwidth]{noncritical-susceptibility-deconfined.eps}
% \end{center} \caption{ susceptibility   }\label{phasedeconfined}
%\end{figure}
%\begin{figure}
%  % Requires \usepackage{graphicx}
%\begin{center}
%\includegraphics[width=.5\textwidth]{noncritical-susceptibility-confined.eps}
% \end{center} \caption{  }\label{phasedeconfined}
%\end{figure}

%%%%%%%%%%%%%%%%%%%%%%%%%%%%%%%%%%%%%%%%%%%%%%%%%%%%%%%%%%%%%%%%%%%%%%%%%%%%%%%%%%%%%%%%%%%%%%%%%%%%%%%%%%%%%%%%
%%%%%%%%%%%%%%%%%%%%%%%%%%%%%%%%%%%%%%%%%%%%%%%%%%%%%%%%%%%%%%%%%%%%%%%%%%%%%%%%%%%%%%%%%%%%%%%%%%%%%%%%%%%%%%
%%%%%%%%%%%%%%%%%%%%%%%%%%%%%%%%%%%%%%%%%%%%%%%%%%%%%%%%%%%%%%%%%%%%%%%%%%%%%%%%%%%%%%%%%%%%%%%%%%%%%%%%%%%%%%%%%%%%%%
%%%%%%%%%%%%%%%%%%%%%%%%%%%%%%%%%%%%%%%%%%%%%%%%%%%%%%%%%%%%%%%%%%%%%%%%%%%%%%%%%%%%%%%%%%%%%%%%%%%%%%%%%%%%%%%%%%%%%%
\section{Kubo formula}
The response of a thermodynamic system to an applied external field is 
described by transport coefficients of the system. For small
deviation from equilibrium, Kubo formula relates  transport
coefficients to the equilibrium retarded  green's functions of the
system. In particular real-time correlator  of two electromagnetism
current determines  electric conductivity of the medium  via

\be\label{kubo}
\sigma=\frac{1}{2}\lim_{\omega\longrightarrow0}\frac{1}{\omega}Im
G^{\alpha \; R}_\alpha(\omega,\textbf{k}) \ee

where $G_{\alpha\alpha}$ is retarded green's function of two
transverse electromagnetic current. In this section we calculate
retarded green's function of two EM currents in the dual QCD by
using of the Lorenzian AdS/CFT prescription\cite{Son:2002sd}.  To
do so, we turn off external electric field, $e=0$, thus the system
is in equilibrium and turn on a nontrivial time-component gauge
field $A_0$  as a background field, corresponding to  a finite
charge density in the dual gauge theory. Therefore the background
is given by

 \bea\label{background} &ds^2|_{D4}&=\bigg(\frac{u}{R_{AdS}}\bigg)^2(-f(u)dt^2+dx^idx^i)+\bigg(\frac{R_{AdS}}{u}\bigg)^2 \frac{du^2}{f(u)}
               \\&F_{tu}&= -A_t'(t,u)\nn\eea

Where $ds^2|_{D4}$ is the  induced metric on
D4-$\overline{\text{D4}}$ flavor branes. According to holographic
dictionary we focus on the  linearized fluctuations of U(1) gauge
field on the gravitational background. By expanding the DBI action
up to second order in the field strength around the background (\ref{background}), we
arrive at

%\be
%S_{D4-D\overline{4}}= {\cal{N}} \int d^5\xi \sqrt{-det(g_{ab}+2\pi \alpha ' F_{ab})}
%\ee
%
%\be
%F_{0u}^{(0)}=-A_t'(t,x)
%\ee

\be\label{kuboaction}\begin{split} S=\frac{(2\pi \alpha')^2 {\cal{N}}}{2R_{AdS}^3}\int \frac{du}{\sqrt{u^6+C^2}}\big{[}
-\frac{(u^6+C^2)}{f(u)}(\frac{R_{AdS}}{u})^4
F_{ti}^2-\frac{(u^6+C^2)^2}{u^6}F_{0u}^2\cr+R_{AdS}^4 u^2
\sum_{i<j}F_{ij}^2+f(u) (u^6+C^2)\sum F_{iu}^2\big{]}\end{split} \ee

where, $C=\frac{-R_{AdS}^3 }{(2\pi\alpha'){\cal{N}}} D$,  it is
convenient  to change  variable to \be y=\frac{R_{AdS}^2}{u} \ee

(\ref{kuboaction}) take the following form

\be\begin{split}\nn S=\frac{(2\pi \alpha')^2 {\cal{N}}R_{AdS}}{2}  \int \frac{dy
y^{-1}}{\sqrt{1+C'^2 y^6}}\big{[} -\frac{(1+C'^2y^6)}{f(y)}
F_{ti}^2-(1+C'^2y^6)F_{0y}^2+\sum_{i<j}F_{ij}^2
\cr+f(y)(1+C'^2y^6)\sum F_{iy}^2\big{]}\end{split} \ee

where $y_T=\frac{R_{AdS}^2}{u_T}=\frac{5}{4\pi T}$ and

\be f(y)=1-\frac{y^5}{y_T^5}     \;\;\;\;\;\;\;\;\;  C'=\frac{C}{R^6} \ee

Since transverse part of electromagnetic current couples to
transverse part of bulk electric field, we consider the e.o.m for
transverse part of electric field. It is convenient to work in
fourier space

\be A_M(x^\mu,y)=\int \frac{d^4k}{(2\pi)^4} e^{ik.x}\,
A_M(k_\mu,y)\;\;\;\;\;\;\;\;\;\;\;\;\;\; k_\mu=(-\omega,q,0,0) \ee

and equation of motion for transverse components, $E_\alpha=\omega
A_\alpha$ ($\alpha=2,3$), becomes

\be\label{eom} E_\alpha
^{''}+\frac{y}{f(y)\sqrt{1+C'^2y^6}}(y^{-1}f(y)\sqrt{1+C'^2 y^6})'
E_\alpha'+\frac{\omega^2-\frac{q^2f(u)}{(1+C'^2
y^6)}}{f(y)^2}E_\alpha=0 \ee

From this equation the near horizon behavior of $E_\alpha$ can be read as

 \be E_\alpha=(y_T-y)^{\pm
iw/2} \ee

where  $\ w=\frac{2\omega y_T}{5}=\frac{\omega}{2\pi T}$ and $\pm$
represent ingoing and outgoing wave into horizon, respectively. By
imposing ingoing boundary condition at the horizon, the behavior of
solution near $y=0$ becomes

\be E_\alpha={\cal{A}}+{\cal{B}}\, y^2 \ee

The relevant part of boundary action for calculating retarded
corellator is

 \bea
S
%=N\int d^{d+1}x \sqrt{-g} g^{AC}g^{BD}F_{AB}F_{CD}\\
&=&\frac{(2\pi \alpha')^2 {\cal{N}}}{2} R_{AdS}\int d^{d+1}x \sqrt{-g} g^{uu}g^{\alpha\alpha}A_{\alpha}\partial _u A_{\alpha}\\
%=-2N\int \frac{dw dq}{(2\pi)^2}\sqrt{-g}g^{uu}g^{\alpha\alpha}\frac{1}{\omega^2}E_{\alpha}'E_{\alpha}\\
&=&\frac{(2\pi \alpha')^2 {\cal{N}}}{2} R_{AdS}\int \frac{dw
dq}{(2\pi)^2}\;\frac{2{\cal{A}} {\cal{B}}}{\omega^2}
 \eea

By applying AdS/CFT reception for calculating real-time correlator,
we have

\be G^{R}_{\alpha\alpha}=\frac{\delta^2S}{\delta A_\alpha
\delta A_\alpha}=\omega^2\frac{\delta^2S}{\delta E_\alpha
\delta E_\alpha}=2(2\pi \alpha')^2 {\cal{N}} R_{AdS}  Im [{\cal{A}}/{\cal{B}}] \ee
%
%\be \chi^\mu_\mu=-4Im C_{\alpha\alpha}=\frac{8RN(2\pi\alpha')^2}{25}
%(4\pi T)^2 Im {\cal{A}}/{\cal{B}} \ee
%
%By plugging   ? into ?   conductivity takes the following form
%
%
%\be \sigma=\frac{1}{2\omega}Im {C^{\alpha}}^{R}_\alpha \ee

In appendix  we have calculated ${\cal{A}}/{\cal{B}}$ at low
frequency, $w\longrightarrow 0$, the result is

\be \frac{{\cal{A}}}{{\cal{B}}}=1+\frac{5}{4} i \sqrt{1+C' y_T^6}\,\frac{w}{y_T^2} \ee

substituting this result into (\ref{kubo})  we arrive at

\be \sigma=\frac{(2\pi \alpha')^2 {\cal{N}}}{ R_{AdS}} u_T
\sqrt{1+\frac{d^2}{u_T^6}} \ee

which is exactly (\ref{conden})  with $e=0$.

\section{Conclusion}
We have studied a noncritical holographic model in external
electric field and compare its results with the ones of
Sakai-Sugimoto model and D3-D7 system. We found  that, remarkably, the behavior
of conductivity as a function of temperature, electric field and
baryon density is similar to the result of D3-D7 system. In
particular in weak-field regime the conductivity grows linear with
temperature, $\sigma\propto T$.

Computation of transport coefficients by using lattice QCD requires an analytic continuation to real-time
space, which leads to systematic errors in results. However, lattice results with small errors provide insights
about strongly coupled regime of QCD. In \cite{Gupta:2003zh} putting systematic errors under control, a lattice simulation
 of conductivity in
QCD has been done, the reported result is

\bea
\frac{\sigma(T)}{T}=C_{EM}\left\{
                      \begin{array}{ll}
                        7.5\pm0.8, \;& \hbox{$T=1.5 T_c$} \\
                        7.7\pm0.6, & \hbox{$T=2  T_c$} \\
                        7.0\pm0.4, & \hbox{$T=3 T_c$}
                      \end{array}
                    \right.
\eea

where electromagnetic vertex factor is given by  $C_{EM}=4\pi\alpha\sum e_f^2$, in which $\alpha$ is fine structure
constant and $e_f$ is electric charge of a quark with flavor f ($C_{EM}\approx \frac{1}{20}$ for two flavors \cite{Gupta:2003zh}).
 If we can trust this result, it shows that, with good accuracy, conductivity is linear in
 temperature, in accordance  with the predictions  of noncritical models and D3-D7 system (soft wall model also predicts a linear
 dependence  $\sigma\propto T$ \cite{Atmaja:2008mt}).

By studding the phase structure  of the model, we observed that the electric field reduces  chiral-restoration temperature.
Also the general structure of phase diagram closely resemble phase diagram of Sakai-Sugimoto model \cite{Bergman:2008sg}. In addition  we have checked our results
by using Kubo formula in section 3.

\section*{Acknowledgment}
The author would like to thank  Mohsen Alishahiha for many helpful
discussions and encouragement and for carefully reading and commenting on the
manuscript. I would also like to thank H.R. Afshar, A. Akhavan, M. Ali-Akbari, D. Allahbakhsi, K. Bitaghsir Fadafan,
R. Fareghbal, A.E. Mosaffa, A. Naseh, A.A  Varshovi, and A.V. Zayakin for discussions.

\section*{Appendix}
In  this appendix we explore the low-frequency behavior of retarded
green's function, by solving (\ref{eom}) perturbatively in $\omega$. Let us
first rewrite (\ref{eom}) in terms of more convenient variable

\be
E_{\perp}=(y_T - y)^{- i w/2} \phi(y)
\ee

substituting this into (\ref{eom}) we  find  the following equation for $\phi(y)$

\bea\label{eom1}
\phi ''(y)+ \mathfrak{P} \,\phi'(y)+\mathfrak{R}\,\phi(y)=0
\eea

%\bea\label{eom1}\begin{split}\nn &\phi ''(y)+ \big{(}\frac{ i w }{1-y}+\frac{
%(1+4 y^5-2 \widetilde{C}  y^6+7 \widetilde{C} y^{11}) }{(-1+y) y
%(1+y+y^2+y^3+y^4) (1+\widetilde{C}
%y^6)}\big{)}\phi'(y)\\&+\big{(}-\frac{w (-2 i+w)
%}{4(-1+y)^2}+\frac{25 w^2 y^5 (1+\widetilde{C}  y) }{4(-1+y^5)^2
%(1+\widetilde{C}  y^6)}\\&
%-\frac{i w (1+4 y^5-2 \widetilde{C} y^6+7
%\widetilde{C} y^{11})}{2(-1+y)^2 y (1+y+y^2+y^3+y^4)
%(1+\widetilde{C} y^6)}\big{)}\phi(y)=0\end{split}\eea

where  we have rescaled variables as $y\longrightarrow y_T\,y $ and $C'\longrightarrow \widetilde{C}=y_T^6\, C'$, for simplicity and

\bea\nn
&\mathfrak{P}&=\frac{ i w }{1-y}+\frac{
(1+4 y^5-2 \widetilde{C}  y^6+7 \widetilde{C} y^{11}) }{(-1+y) y
(1+y+y^2+y^3+y^4) (1+\widetilde{C}
y^6)}\\\nn
&\mathfrak{R}&=-\frac{w (-2 i+w)
}{4(-1+y)^2}+\frac{25 w^2 y^5 (1+\widetilde{C}  y) }{4(-1+y^5)^2
(1+\widetilde{C}  y^6)}
-\frac{i w (1+4 y^5-2 \widetilde{C} y^6+7
\widetilde{C} y^{11})}{2(-1+y)^2 y (1+y+y^2+y^3+y^4)
(1+\widetilde{C} y^6)}
\eea

In order to solve this equation  perturbatively in $w$, we consider following expansion for $\phi$

\be \phi=\phi_0+\omega \phi_1 \ee

plugging this into (\ref{eom1}) and imposing regularity condition at the
horizon, we find that $\phi_0$ is a constant and $\phi_1$ satisfies
following equation

\be \phi_1 ''(y)+P(y)\phi_1 '(y)=R(y) \ee where

\bea
&P(y)&=\frac{1+4 y^5-2 y^6 \widetilde{C}  +7 y^{11} \widetilde{C}  }{(-y+y^6) (1+y^6 \widetilde{C})}\\
&R(y)&=i  [-\frac{1}{2 (-1+y)^2} +\frac{1+4 y^5-2 y^6 \widetilde{C}
+7 y^{11} \widetilde{C}  }{2 (-1+y)^2 y (1+y+y^2+y^3+y^4) (1+y^6
\widetilde{C})}]\eea

The general solution of above equation is
\be\label{sol}
\phi_1(y)=\lambda \widetilde{\phi}(y)+\widetilde{\phi}(y)\int_0^y dx \frac{R(x)}{\widetilde{\phi}'(x)}-\int_0^y dx \frac{R(x)
\widetilde{\phi}(x)}{\widetilde{\phi}'(x)}
\ee
where $\lambda$ is a constant and  particular  solution, $\widetilde{\phi}(y)$,  is given by

\be
\widetilde{\phi}(y)=\int_0^y dx e^{-\int_0^x dz P(z)}
\ee

For our purpose, it is sufficient to consider the near horizon and near boundary form of the solution. First consider the near horizon behavior
of solution

\be \phi(y)\simeq  \frac{\lambda}{5 \sqrt{1+\widetilde{C}} } \ln(y-1)+\frac{iw}{2}\ln(y-1) \ee

regularity at horizon implies that
\be\label{lambda}
\lambda=-\frac{5iw}{2}\sqrt{1+\widetilde{C}}
\ee

Take the limit $y\longrightarrow 0$ in  (\ref{sol})and using (\ref{lambda}), we  find the asymptotic form of solution
 near the boundary as
\be
\phi(y)\simeq 1-i w (-5 \sqrt{1+\widetilde{C}}\frac{y^2}{4}+\frac{y^2}{4}+\frac{y}{2})
\ee

and in original coordinate we have

\be
E_{\alpha}(y)\mid_{y\longrightarrow 0} \simeq 1+\frac{5i\, w }{4} \sqrt{1+C'\, y_T^6}\; y^2/y_T^2
\ee

\end{document}